\documentstyle[aps,preprint,tighten,epsf]{revtex}
\def \bfgr #1{ \mbox {{\boldmath $#1$}}}
\newcommand{\be}{\begin{eqnarray}&&}
\newcommand {\CM} {{\cal M}}
\newcommand{\ee}{\end{eqnarray}}
\newcommand{\la}{\langle\,}
\newcommand{\ra}{\,\rangle}
\def\bp{{\mbox{\boldmath$p$}}}
\def\bn{{\mbox{\boldmath$n$}}}

\def\bxi{{\mbox{\boldmath$\xi$}}}
\def\bsigma{{\mbox{\boldmath$\sigma$}}}
\begin{document}

\title{Polarization observables in the reaction
\mbox{\boldmath$p n \to d \phi$}
}
\author{
{\sc L.P. Kaptari}\thanks{On leave of absence from Bogoliubov Laboratory
of Theoretical Physics, JINR, 141980, Dubna, Moscow reg., Russia },
{\sc B. K\"ampfer}}

\address{Forschungszentrum Rossendorf, PF 510119, 01314 Dresden, Germany}
\maketitle

\begin{abstract}
The reaction  $p n \to d \phi$ is studied within a covariant boson
exchange model. The behavior of polarization observables
being accessible in forthcoming experiments near threshold
is predicted.\\[3mm]
PACS. 13.75.-n Hadron-induced low- and intermediate-energy reactions
and scattering (energy $\le$ 10 GeV)
--
14.20.-c Baryons (including antiparticles)
-- 
21.45.+v Few-body systems
\end{abstract}
 
\section{Introduction}

Data on elementary reactions with neutrons are scare since 
either they must be
extracted, with some efforts and even mostly with some model dependent
assumptions, from reactions on nuclei,
or
a tagged neutron beam (cf.\ \cite{IUCF}) is used. 
The spectator technique
\cite{johansson,anke}
represents one example how one can use a deuteron target to isolate
quasi-free reactions at the neutron. It is based on the idea to measure the
spectator proton ($p_{\rm sp}$) at fixed
(or slightly varying) beam energy in the
meson ($M$) production reactions $p d \to d M p_{\rm sp}$
thus exploiting the internal momentum spread of the neutron inside the
deuteron ($d$). In such a way one gets access to quasi-free reactions
$p n \to d M$ if, in experiments with the deuteron target at rest,
the spectator proton has momenta in the order of 50 $\cdots$ 150 MeV/c.

An experimental investigation of the threshold-near
(pseudo)scalar and vector meson production at the neutron
becomes therefore feasible. Indeed, at COSY the ANKE spectrometer
set-up can be used, in particular, for studying the $a_0$, $\omega$ and
$\phi$ production with the internal beam at a ''neutron target'' \cite{anke}.
This offers the possibility to enlarge the data base on hadronic
reactions and to address special issues.
For instance, there is already a large body of data which can be used
for a systematic study of the OZI rule violation via $\omega$ and
$\phi$ production in $\pi N$ and $p p$ reactions
(cf.\ \cite{sibirtsev} for a reanalysis) and in $\bar p p$ annihilation
(cf.\ \cite{ellis,rotz} for theoretical
analyzes) as well. OZI rule violations are of interest with respect
to possible hints to a significant $s \bar s$ admixture in the proton,
as supported by the pion-nucleon $\Sigma$ term \cite{donoghue,gasser}
and interpretations of the lepton deep-inelastic scattering data
\cite{ashman}.
Besides the impact on hadron phenomenology
the origin of the OZI rule has
also a link to QCD \cite{isgur,shuryak}.

In the reactions $pp \to pp M$ or $pn \to pn M$ the final state interaction
among the outgoing nucleons plays a role.  Therefore, the meson production
process is a convolution of the ''pure production process'' and the final
state interaction. In the case of the reaction $pn \to d M$ one has
{\em one} well defined final state of the nucleons and may better constrain
the elementary production amplitude.

Finally, we mention that in the reactions
$pp \to pp M$, $pn \to pn M$ and $pn \to d M$ the conservation laws and
symmetry principles determine a different dynamics near threshold, which
needs to be investigated to allow a firm understanding of the
reactions and the systematics of the OZI rule violation.

Given this motivation, in \cite{grishina_omega,nakayama}
the reaction $p n \to d V$ with
$V = \omega, \phi$ has been studied in some detail.
($p n \to d S$ with $S = a_0^+, \eta, \eta'$ is considered
in \cite{grishina_scalar}.)
In \cite{grishina_omega} the cross sections and angular
distributions are elaborated as
a function of the excess energy within a  two-step model.
The same observables are evaluated in \cite{nakayama}
within the framework
of a boson exchange model with emphasis on the ratio of cross sections
$\sigma_{pn \to d \phi} / \sigma_{pn \to d \omega}$
being of direct relevance for the OZI rule violation.

Since at COSY the proton beam is polarized and also polarized targets
are envisaged, we extend the previous studies \cite{grishina_omega,nakayama}
to make a prediction of polarization observables
in the reaction $p n \to d \phi$.
The set-up of the ANKE experiment
at COSY can directly identify the $\phi$ via its $K^+ K^-$ decay
channel.  
We present the asymmetry,
tensor analyzing power, and
proton-phi spin-spin correlations.
In doing so we use our previous one-boson exchange model \cite{titov}
with parameters adjusted to available threshold-near data
on $\phi$ production in $pp$ and $\pi p$ reactions and combine this
with our previous studies \cite{titov,ourphysrev,quad} employing
a solution of the deuteron wave function within the
Bethe-Salpeter (BS) formalism. In this way we derive a completely covariant
approach with correct treatment of the off-shell effects in the
$\pi N \to \phi N$ subprocess amplitudes.

Extensions to $p n \to d \omega$ and
even to $p d \to d V p_{sp}$ 
are much more involved and, therefore, relegated to separate work.

Our paper is organized as follows. In section 2 we present the
theoretical framework and elaborate the basic equations. 
The numerical results and their discussion are presented in section 3.
The summary can be found in section 4.

\section{The model}

\begin{figure}
\epsfxsize 2.8in
\centerline{\epsfbox{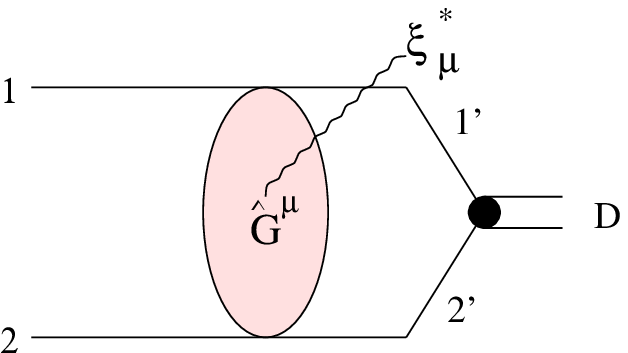}}
~\vskip 3mm
\caption{ The diagram for the process 
$p(1) + n(2) = d + \phi$.
$G$ is the scattering operator, and $\xi^*$ denotes
the polarization vector of the
outgoing  vector meson.}
\label{diagr1}
\end{figure}

The invariant differential cross section 
of the reaction $p n \to d \phi$ reads
\be
\frac{d\sigma}{dt} = \frac{1}{16\pi\,s(s-4m^2)}\,\frac{1}{4}\sum\limits_{s_1,s_2}
\sum\limits_{\CM_\phi,\CM_d}
\,|T_{s_1s_2}^{\CM_\phi\CM_d}(s,t)|^2,
\label{eq1}
\ee
where $s$ is the square of the total energy of the colliding particles 
p and n
in the center of mass, $t$ is the square of the
transfered 4-momentum, $m$ is the nucleonic mass,
$s_1,s_2,\CM_\phi$ and $\CM_d$ denote the spin projections
on a given quantization axis, and
$T$ stands for the invariant amplitude.
The general form of $T$
may be written in the form (cf.\ Fig.~\ref{diagr1})
$T_{s_1s_2}^{\CM_\phi\CM_d}(s,t)
=
\la d,\CM_d\,|\hat G_\mu\,\xi^{*\mu}_{\CM_\phi}|1,2\ra, $
where $\xi^\mu_{\CM_\phi}$ is the polarization 4-vector
of the $\phi$ meson.  The scattering
operator $\,\hat G\,\,$ represents a 4-vector in Minkowski space,
a vector in the vector space of mesons and a
$16\otimes 16$ component object in the spinor space of nucleons.
The deuteron is
described as a $16$ component BS amplitude $\Phi(1,2)$ which
is defined as a matrix element of a time ordered product of two nucleon
fields $\psi(x)$ by
$
\Phi^{\alpha\beta}(1,2)=\la d| {\cal T}
\left(\psi^\alpha(1)\psi^\beta(2)\right)|0\ra
$
and satisfies the BS equation.
By defining another scattering operator 
via
$\hat O = \hat G_\mu \, \xi^{\mu *}_{\CM_\phi}$
the invariant amplitude reads
\be
T_{s_1s_2}^{\CM_\phi\CM_D}(s,t)
=
-i \int \, \frac{d^4p}{(2\pi)^4} \,
\bar \Phi_{\CM_d}^{\alpha b}(1',2') \,
\hat O^{bc}_{\alpha\beta}(12;1'2',\CM_\phi)
\, u^c_{s_1}(1) \, u^\beta_{s_2}(2),
\label{eq5}
\ee
where $\bar\Phi_{\CM_d}^{\alpha b}(1',2')$ is the conjugate
BS amplitude in the momentum space,
$p$ is the relative 4-momentum of the nucleons in the deuteron, and
$u(1)$ and $u(2)$ denote the Dirac spinors for the incident nucleons.  
Summation over spin indices
$\alpha, \beta, b, c = 1 \cdots 4$ occurring pairwise is supposed.
The operator $\hat O^{bc}_{\alpha\beta}(12;1'2',\CM_\phi)$
is a scattering operator describing
the $\phi$ meson production in the final state.
This operator acts in the spinor space
of protons and neutrons separately; the upper (Latin)
and lower (Greek) spinor indices
refer to protons and neutrons, respectively.
The first indices, $b$ and $\alpha$,  form
an outer product of two columns, whereas the second ones, $c$ and $\beta$,
form an outer product of two rows.
To specify explicitly the spinor structure
we decompose the operator $\hat O$ in each of its indices
over the corresponding complete set of Dirac spinors, i.e,
\be
\hat O^{bc}_{\alpha\beta}(12;1'2',\CM_\phi)
=
\frac{1}{(2m)^4}\sum\limits_{r,r',\rho,\rho'=1}^4\,
A_{r r', \rho \rho'}^{\CM_\phi}(12;1'2')
 u^b_{r'}(1') \bar u^c_{r}(1)  \bar u^\beta_{\rho}(2)  u^\alpha_{\rho'}(2'),
\label{eq6}
\ee
where the coefficients $A_{r r', \rho \rho'}^{\CM_\phi}(12;1'2')$  
may be found by
using the completeness and orthogonality of the Dirac spinors,
$\bar u_r({\bf p}) u_{r'}({\bf p})
=
2 \varepsilon_r m \delta_{rr'}$, yielding
\be
A_{r r' , \rho \rho'}^{\CM_\phi}(12;1'2')
=
\varepsilon_r \varepsilon_{r'} \varepsilon_\rho \varepsilon_{\rho'} \,
\bar u^b_{r'}(1') \, 
\bar u^\alpha_{\rho'}(2') \, 
\hat O^{bc}_{\alpha\beta}(12;1'2',\CM_\phi) \,
u^c_{r}(1) \, u^\beta_{\rho}(2),
\label{eq7}
\ee
where $\varepsilon_r = +1$ for  $r=1,2$ and
$\varepsilon_r = -1$ for $r=3,4$.
Substituting (\ref{eq6}) into (\ref{eq5}) one obtains
\be
T_{s_1s_2}^{\CM_\phi\CM_d}(s,t)
=
\frac{-i}{(2m)^2} \int \, \frac{d^4p}{(2\pi)^4}
\bar \Phi_{\CM_D}^{\alpha b}(1',2') \,
\sum\limits_{r, r' = 1}^2 A_{s_1 s_2, r r'}^{\CM_\phi}(12;1'2') \,
u^\alpha_{r'}(2')u^b_r(1')\nonumber\\ &&
=
\frac{i}{(2m)^2}\sum\limits_{r, r'=1}^2
\int\,\frac{d^4p}{(2\pi)^4}
\left( u^\alpha_{r'}(2')\right )^T\gamma_c^{\alpha\alpha'}
\left( \gamma_c^{\alpha'\alpha''}
\bar \Phi_{\CM_d}^{\alpha'' b}(1',2')\right )\,
A_{s_1 s_2, r r'}^{\CM_\phi}(12;1'2')
u^b_r(1')\nonumber\\&& =
\frac{i}{(2m)^2}\sum\limits_{rr'=1}^2
\int\,\frac{d^4p}{(2\pi)^4} A_{s_1 s_2, r r'}^{\CM_\phi}(12;1'2')
\bar { v}_{r'}(2') \bar\Psi_{\CM_d}(1',2')u_r(1'),
\label{eq8}
\ee
where $\gamma_c$ is the charge conjugation matrix,
$\bar v_r(2') \equiv (u_r(2') )^T \gamma_c$, and
the new BS amplitude
$\bar \Psi_{\CM_d}(1',2')\equiv \gamma_c\bar\Phi_{\CM_d}(1',2')$
now is a $4\otimes 4$ matrix
and represents the solution of the BS
equation written also in  matrix form.

\begin{figure}
\epsfxsize 6.5in
\epsfbox{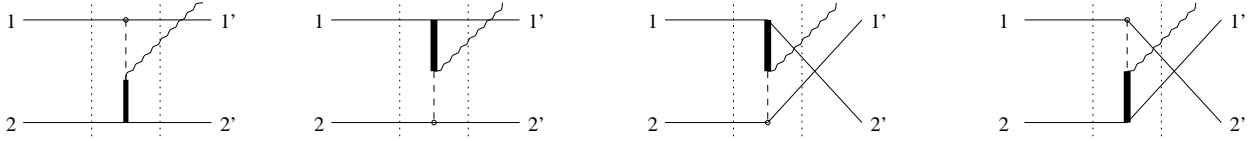}
~\vskip 3mm
\caption{ Graphical representation of the operator $\hat O$
defined in (\ref{eq6}) in the
one-boson exchange approximation.
Nucleons in the initial and
final states are represented by thin lines which are
truncated by vertical dotted lines as to obtain an operator.
The exchanged $\pi$ and $\rho$ mesons are depicted by
vertical dashed and thick lines, respectively. The four
different contributions correspond to different combinations
of $\pi^0,\pi^\pm$ and $\rho^0,\rho^\pm$ exchanges.}
\label{diagr2}
\end{figure}

For further evaluations of the amplitude
(\ref{eq8}) one needs to specify an explicit form of the operator $\hat O$.
Following \cite{titov},
and taking into account only the meson-exchange contribution from
one-boson exchange diagrams with internal meson conversion, 
the operator $\hat O$ may be
represented by four truncated diagrams as depicted in Fig.~\ref{diagr2}.
(The contributions of the nucleonic currents
are found  to be negligibly small
\cite{nakayama,titov} and, therefore, can be omitted for many purposes.)
Consequently, having computed the operator $\hat O$
by using the effective interaction Lagrangians for the
$\pi NN$, $\rho NN$, $\phi \rho \pi$ vertices 
and monopole form factors for dressing the vertices as in \cite{titov},
it is straightforward to obtain the coefficients
$A_{r r', \rho \rho'}^{\CM_\phi}(12;1'2')$
in (\ref{eq7}).
If all particles were on mass shell,
$A_{r r' , \rho \rho'}^{\CM_\phi}(12;1'2')$
exactly coincides with the amplitude of the elementary
process $1 + 2 \to 1' + 2' + \phi$. 
However, in general case this amplitude corresponds to
a virtual process of vector meson production
with two off-shell nucleons in the final state.

Since our numerical solution \cite{solution} of
the BS equation 
has been obtained in the deuteron's center of mass,
all further calculations will be performed in this system.
First, as depicted in Fig.~\ref{diagr1},
we introduce the relevant kinematical variables as follows:
$p_{1,2}$ are the 4-momenta of incoming nucleons,
$p_{1,2}'$ stand for the 4-momenta
of the internal (off-shell) nucleons in the deuteron with
$p=(p_1'-p_2')/2$;
$\xi_{\CM_d}$ denotes the polarization 4-vector of the deuteron.
In this notation the BS amplitudes in the
deuteron's rest system are of the form \cite{quad}
\begin{eqnarray}
&&
\!\!\!\!\!
\Psi_{\CM_d}^{S^{++}}(p_1',p_2')
=
{\cal N}(\hat k_1+m )\frac{1+\gamma_0}{2}\hat\xi_{\CM_d}(\hat k_2-m)
\phi_S (p_0,|{\bf p}|), \label{psis}\\[2mm]
&&\!\!\!\!\!
\Psi_{\CM_d}^{D^{++}}(p_1',p_2')=-\frac{{\cal N}}{\sqrt{2}}
(\hat k_1+m )\frac{1+\gamma_0}{2}
\left (
\hat\xi_{\CM_d} +\frac{3}{2|{\bf p}|^2} (\hat k_1-\hat k_2)(p\xi_M)\right )
(\hat k_2-m)
\phi_D (p_0,|{\bf p}|),\nonumber
\end{eqnarray}
where $\, \hat{} \,$ means contraction with Dirac matrices,
and
$k_{1,2}$ are on-shell 4-vectors related  to the
off-shell vectors $p_{1,2}'$ as follows
\begin{equation}
k_1=(E_p,\bp),\quad k_2=(E_p,-\bp),\quad p_1'=(p_{10}',\bp),\quad
p_2'=(p_{20}',-\bp),\quad E_p=\sqrt{\bp^2+m^2},
\nonumber
\end{equation}
and $\phi_{S,D} (p_0,|{\bf p}|)$ are the partial scalar amplitudes, related to
the corresponding partial vertices as
\begin{equation}
\phi_{S,D} (p_0,|{\bf p}|)=
\displaystyle\frac{G_{S,D} (p_0,|{\bf p}|)}{\left(
\displaystyle\frac12 M_d - E_p \right)^2-p_0^2}.
\nonumber
\end{equation}
$M_d$ is the deuteron mass, and the normalization factor is
${\cal N} =\left\{ \sqrt{8\pi} 2E (E+m) \right\}^{-1}$.
To be explicit let us recall
the components of the polarization 4-vector of a vector particle with
4-momentum  $p=(E,{\bf p})$, polarization index $\CM=\pm 1,\, 0$ 
and mass $M$ as
\begin{eqnarray}
&&
\xi_\CM=\left( \frac {\mbox{\boldmath{$p$}}
\mbox{\boldmath{$\xi$}}_\CM}{M},
\mbox{\boldmath{$\xi$}}_\CM+\mbox{\boldmath{$p$}}
\frac {\mbox{\boldmath{$p$}} \mbox{\boldmath{$\xi$}}_\CM}{M(E+M)} \right ),
\label{xi}
\end{eqnarray}
where $\bxi_\CM$ is the polarization 3-vector for the particle at rest with
$\bxi_{+ 1} = \frac{1}{\sqrt{2}}(1,  i, 0)$,
$\bxi_{- 1} = \frac{1}{\sqrt{2}}(1, -i, 0)$,
$\bxi_{  0} = (0, 0, 1)$
and
the above Dirac spinors, normalized  as $\bar{u}(p) u(p)=2m$
and $\bar{v}(p) v(p)=-2m$, read
\begin{eqnarray}
u(\bp,s)=\sqrt{m+\epsilon}
\left ( \begin{array}{c} \chi_s \\ \frac{\bsigma \bp}{m+\epsilon}\chi_s
\end{array}  \right), \quad \quad
v(\bp\,,s)=\sqrt{m+\epsilon}\left( \begin{array}{c}
\frac{\bsigma \bp}{m+\epsilon}\widetilde\chi_{s}\\
 \widetilde\chi_{s}
\end{array}  \right),
\label{spinors}
\end{eqnarray}
where $\widetilde\chi_s\equiv -i\sigma_y\chi_s$,  and
$\chi_s$ denotes the usual two-dimensional Pauli spinor.
In general, the BS amplitude consists on eight partial components.
In (\ref{psis}) we take into account only 
the most important ones, namely the
$S$ and $D$ partial amplitudes. The other six amplitudes may become important
at high transferred momenta \cite{ourphysrev,quad}, hence for the present
near-threshold process they may be safely disregarded.
Substituting (\ref{psis}-\ref{spinors}) into
(\ref{eq8})
one obtains after some algebra
\begin{eqnarray}
&&T_{s_1s_2}^{\CM_\phi\CM_d}(s,t)
=
\frac{-i}{\sqrt{8\pi}}\sqrt{|\CM_D|+1}\label{eq15} \\
&& \times
\sum\limits_{r,r'}
\int\,\frac{d^4p}{2E_p(2\pi)^4}
\frac{G_S-G_D \frac{1}{\sqrt{2}}}{\left(\frac12 M_d - E_p \right)^2
-p_0^2} A_{s_1 s_2, r r'}^{\CM_\phi}
(\bp_1 \bp_2, \bp_1' \bp_2')
\, \delta_{r+r',\CM_d} \nonumber \\
&& +
\frac{3i}{\sqrt{16\pi}}
\sum\limits_{r,r'}
\int\,\frac{d^4p}{2E_p(2\pi)^4}
\frac{G_D}{\left(\frac 12 M_d - E_p\right)^2-p_0^2}
A_{s_1 s_2, r r'}^{\CM_\phi}
(\bp_1 \bp_2, \bp_1'\bp_2')
\widetilde\chi^+_{r'}\, (\bfgr\sigma\bn)\chi_r\,(\bn\bxi^*_{\CM_D}).\nonumber
\end{eqnarray}
where here and below the sums over $r, r'$ run from 1 to 2.
By closing the integration contour in the upper hemisphere and
picking up the residuum at $p_0 = M_d/2 - E_p$ and introducing the
notion of the deuteron $S$ and $D$ wave functions as
\be
u_S(p)=\frac{G_S(p_0,|\bp|)}{4 \pi \sqrt{2M_d}(2E_p-M_d)},
\quad
u_D(p)=\frac{G_D(p_0,|\bp|)}{4 \pi \sqrt{2M_d}(2E_p-M_d)}
\label{norm}
\ee
with $2\int d\,|\bp| \, |\bp |^2 (u_S^2+u_D^2) \approx \pi$,
the final expression for the amplitude may be written as
\be
T_{s_1s_2}^{\CM_\phi\CM_d}(s,t) = \sqrt{\frac{M_d}{4\pi}}
\sum\limits_{rr'}
\int\,\frac{d^3\bp}{E_p(2\pi)^2}
A_{s_1 s_2, r r'}^{\CM_\phi}
(\bp_1, \bp_2; \bp, -\bp)
\nonumber\\&&\times
\left\{ \sqrt{|\CM_d|+1} 
\left[
u_S(p) - \frac{u_D(p)}{\sqrt{2}} 
\right]
\delta_{r+r',\CM_d} -
3 \frac{u_D(p)}{\sqrt{2}}
\left( \bxi^*_{\CM_d}\bn \right) \, 
\widetilde \chi_{r'}^+ (\bfgr\sigma\bn)\chi_r
\right\},
\label{finalT}
\ee
where $\bn$ is a unit vector parallel to $\bp$.

\section{Discussion of results}

\begin{figure}
\epsfxsize 3.8in
\centerline{\epsfbox{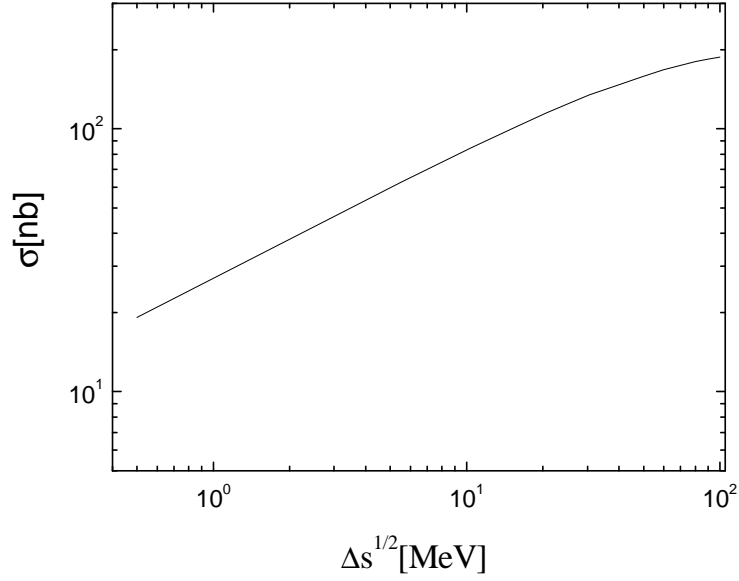}}
~\vskip 3mm
\caption{Total cross section for the reaction $p n \to d \phi$
as a function  of the excess energy
$\Delta s^{1/2}=\sqrt{s}-M_d-M_\phi$.}
\label{Xtotal}
\end{figure}

In our evaluation of the above equations
we use, for describing the deuteron wave function \cite{solution},
the numerical solution of the BS equation in ladder
approximation obtained with a realistic
one-boson exchange interaction which includes
$\pi,\sigma,\omega,\delta,\rho$ and $\eta$ exchanges.
The effective parameters used in the ladder approximation
have been fixed in such a way  to obtain
a good description of the $NN$ elastic scattering data and
the static properties of the deuteron \cite{quad}.
Independent of this set of parameters related to deuteron properties,
the effective coupling constants 
and the attributed cut-off factors
are taken from the recent analysis \cite{titov}.
In all subsequent numerical calculations,
the set B from \cite{titov} is used.
For this parameter set the meson exchange term is by far
dominating. 
In Fig.~\ref{Xtotal} the total cross section
near the threshold is depicted.
The shape of our cross section is rather similar to the one computed in
\cite{nakayama}. However, there is a difference in the absolute values
by roughly a factor of $2/3$.
This difference is in the same order of magnitude as the difference
of the previously often used cross section $\sigma = 0.26$ $\mu$b 
deduced from \cite{DISTO1} and the published value $\sigma = 0.19$ $\mu$b
\cite{DISTO2} for the reaction $p p \to p p \phi$
at an excess energy of 83 MeV.

\begin{figure}
\epsfxsize 3in
\centerline{\epsfbox{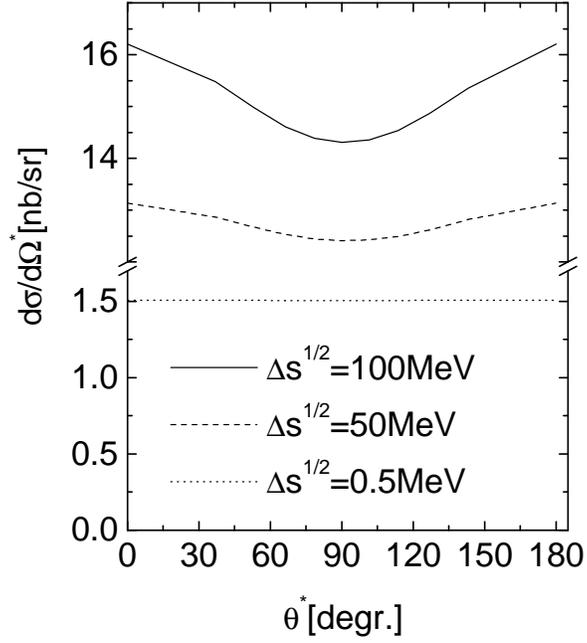}}
~\vskip 3mm
\caption{Angular distribution in the center of mass system
for various values of the excess energy.}
\label{angular}
\end{figure}

Refs.~\cite{nakayama,titov} show that
several sets of parameters equally well describe the $pp\to pp\phi$ data.
These sets differ not only by absolute values of parameters
but also by the relative contributions of meson current and
the nucleon current terms. Since in case of the
$p n \to d \phi$ processes the isospin transition corresponds to
$\Delta I = 0$
the meson-exchange diagrams are enhanced by a factor of three
in comparison with the nucleon current terms.
This means that
(i) the contributions of the nucleon current are suppressed by
about one order of magnitude in comparison with the meson current, and
(ii) the behavior of the cross section
and angular distribution in the process $ p n \to d \phi$
is expected to follow essentially the
behavior of the dominating meson-exchange current contribution 
in the elementary processes $p n \to p n \phi$.
The occurrence of the deuteron wave function will only modify this behavior.
This is clearly seen in Fig.~\ref{angular}, where the behavior of the
angular distribution is very similar to the distribution
found in \cite{titov} for the reaction $p n \to p n \phi$.
At the threshold the distribution is fairly flat,
while with increasing excess energy some
forward-backward peaking becomes visible.
This feature depends upon the parameter set used.
For instance, if we  were using the set C of \cite{titov},
then a slight suppression of the cross section would be
expected in the forward-backward directions as
predicted for the elementary cross section \cite{titov}.
This sensitivity of the predicted angular distribution has been pointed out
in \cite{nakayama,juelich} as a tool to constrain further the parameters,
if precise data become available.

\begin{figure}
\epsfxsize 3.3in
\centerline{\epsfbox{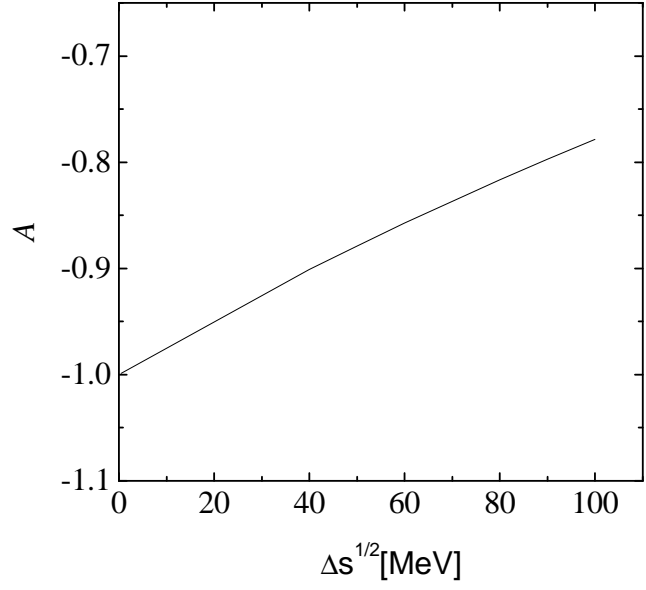}}
~\vskip 3mm
\caption{The beam-target asymmetry ${\cal A}$ 
as a function of the  the excess energy.}
\label{assym}
\end{figure}

Let us now discuss the polarization observables.
Near the threshold, in the final state one has: $I_f = 0$,
$L_f = 0,2$,
$J_f^\pi = J_\phi^\pi + J_D^\pi = 0^-,1^-,2^-$,
where $I$, $L$ and $J^\pi$ are
the total isospin, total radial angular momentum, angular momentum  and parity,
respectively.  Thus from symmetry constraints, in the
initial state the allowed configurations are $I_i = 0$,
$L_i = 1,3,5 \cdots$ and total spin $S_i = 0$.
The conservation law implies $L_i=1$, so that $J_i^\pi=1^-$.
From these transitions
in $|T_{s_1s_2}^{\CM_\phi\CM_d}(s,t)|^2$ one may form
different combinations of spin observables, which near the threshold  behave
 quite differently. If one considers, for example, the cross section
averaged over all final projections $\CM$ at different initial
spin projections, then the beam-target asymmetry, defined here as
\be
{\cal A}=\frac{\sigma(s_p+s_n=1)+\sigma(s_p+s_n=-1)-\sigma(s_p+s_n=0)}
{\sigma(s_p+s_n=1)+\sigma(s_p+s_n=-1)+\sigma(s_p+s_n=0)},
\label{assymetry}
\ee
is predicted to be $-1$ near threshold and to increase with increasing
energy. This is in contrast with the  asymmetry defined for
the elementary process $pp\to pp\phi$ which is $+1$,
as predicted in \cite{titov}, and decreases with increasing energy.
The energy dependence of the asymmetry (\ref{assymetry})
is depicted in Fig.~\ref{assym}.

\begin{figure}
\epsfxsize 3.3in
\centerline{\epsfbox{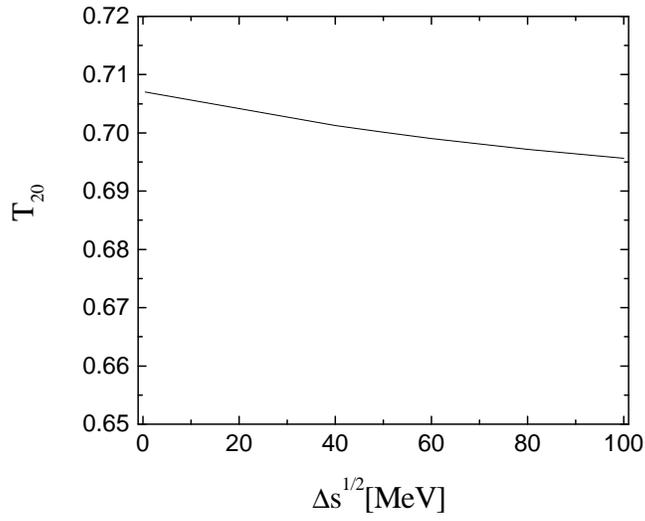}}
~\vskip 3mm
\caption{Deuteron tensor analyzing power $T_{20}$
as a function of the excess energy.}
\label{t20fig}
\end{figure}

Another interesting polarization observable is the tensor analyzing power,
which may be defined either for the final deuteron or for the final meson.
For instance, the deuteron
tensor analyzing power is defined by the cross section
averaged  over all spins but the deuteron,
\be
T_{20}=
\frac{1}{\sqrt{2}}\frac{\sigma(\CM_d=1)+\sigma(\CM_d=-1)-2\sigma(\CM_d=0)}
{\sigma(\CM_d=1)+\sigma(\CM_d=-1)+\sigma(\CM_d=0)}.
\label{t20}
\ee
It is seen from (\ref{t20}) that in line with 
the above discussed selection rules,
$\sigma(\CM_d=0)\ll \sigma(\CM_d=\pm 1)$ and
the deuteron tensor analyzing power is predicted to be almost constant,
$T_{20}\approx 1 / \sqrt{2}$, in a large
region of the energy excess. 
This prediction is illustrated in Fig.~\ref{t20fig}.

\begin{figure}
\epsfxsize 3.5in
\centerline{\epsfbox{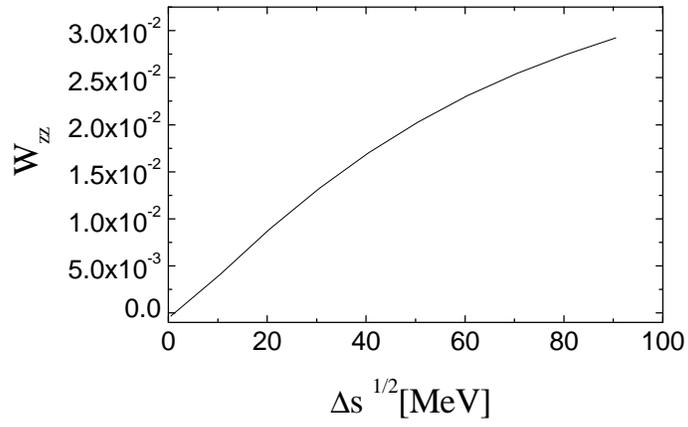}}
~\vskip 3mm
\caption{Proton-phi spin-spin correlation $W_{zz}$
as a function of the excess energy.}
\label{Wzz}
\end{figure}

In Fig.~\ref{Wzz} another polarization observable is depicted
defined by
 \be
 W_{zz}=\left .\frac{\sigma(\CM_\phi=+1)- \sigma(\CM_\phi=-1)}
 {\sigma(\CM_\phi=+1)+ \sigma(\CM_\phi=-1)} \right|_{s_p=+1/2}
 \label{correl}
 \ee
 which characterizes the proton-meson spin-spin correlation. This quantity
 is predicted to vanish near the threshold for the meson exchange diagrams.
 A substantial deviation of $W_{zz}$ from zero would 
 point to the presence of non-negligible nucleon current contributions
 in the $\phi$ production process. In
 Fig.~\ref{Wzz} it is seen that there is some weak dependence
 of $W_{zz}$ upon the energy excess.
 Nevertheless $W_{zz}$ remains very small
 indicating that the polarizations of the incident proton and
 outgoing meson are almost uncorrelated.
 A different situation will occur in the reaction 
$p d \to d\phi p_{sp}$. In
 this case, if one measures also the polarization of the spectator proton,
 the quantity $W_{zz}$ is expected to strongly depend on whether the polarizations
 of protons are the same or have opposite directions.

 Since we have $T_{s_1s_2}^{\CM_\phi\CM_d}(s,t)$ at disposal,
 via (\ref{eq1}) any other polarization observable is accessible
 within the our formalism and corresponding numerical code.

\section{Summary}

In summary we present here a evaluation
of polarization observables for the process $p n \to d \phi$
which are accessible in forthcoming experiments.
Off-shell effects in the sub-process $\pi N \to \phi N$
can correctly be dealt with if one restricts the treatment on the by far
dominating meson-exchange current. It is the beam-target asymmetry
which differs drastically from the one in the reaction
$p p  \to p p \phi$. The tensor analyzing power is fairly insensitive
to variations of the excess energy,
and the proton-phi spin-spin correlation is very small.

The treatment can be extended to incorporate the $\omega$ meson production;
work along this line is in progress.

\subsection*{Acknowledgments}

We are grateful to A.I. Titov for many valuable discussions.
The work is supported in parts by BMBF grant 06DR921 
and the Landau-Heisenberg program.

\newpage

\vspace*{6mm}

\newpage

\end{document}